\documentclass{aa}  

\usepackage{graphicx}
\usepackage{color}
\usepackage{amsmath}
\usepackage{subcaption}
\usepackage{txfonts}
\usepackage{xcolor}
\usepackage[dvipsnames]{xcolor}

\newcommand{\St}{\mathrm{St}}

\begin{document}

\title{Dust dynamics in disk dust traps and \\ late planetesimal formation}
\author{  Tatarelli, M.\inst{1}, Morbidelli, A.\inst{1,2} and Lega, E.\inst{1}}
\institute{ Université Côte d'Azur, Observatoire de la Côte d'Azur, CNRS, Laboratoire Lagrange, France\label{inst1}
   \and
    Collège de France, 11 Pl. Berthelot, 75005 Paris, France \label{inst2}
}

\abstract 
{The streaming instability (SI) is currently the leading model for planetesimal formation in protoplanetary disks, but it typically operates on the timescales of the radial drift time of solids toward the star, i.e. the first $\sim$ Myr. In the Solar System, some planetesimals (the chondrite parent bodies) formed 2–4 Myr after disk formation, implying that dust must have been retained in the disk for extended periods. Pressure bumps provide an efficient mechanism for trapping dust. However, dust trapping alone does not guarantee planetesimal formation: even modest levels of gas turbulence can inhibit strong vertical settling and radial concentration, preventing the dust density from reaching the threshold required for gravitational collapse. This motivates the exploration of alternative dust–gas instabilities, such as the Dusty Rossby Wave Instability (DRWI), which has first been studied in 2D shearing-box simulations.}
{We aim to investigate the viability of such alternative instabilities in global disk simulations under realistic physical conditions.}
{We use the numerical code fargOCA, in which the treatment of dust as a pressureless fluid has been recently implemented. We first recover the results of previous 2D shearing-box simulations using global 2D disk simulations, and then extend the analysis to fully 3D disks in both viscous and inviscid regimes.}
{We reproduce previous 2D results and extend them by characterizing the dust clumping produced by the DRWI in a viscous disk ($\alpha=10^{-4}$). We find that this instability does not develop in fully 3D viscous disks, quenched by high-z gas layers that remain unperturbed due to the settling of dust near the midplane. Motivated by this suppression, we explore the inviscid limit and find that multiple dust sub-rings form, concentrating solids into several thin ring-like structures. These structures would not be resolved in observations and would appear as a single radially broad and vertically thin ring. This explains the geometry of the rings observed in protoplanetary disks without the need to invoke anisotropic turbulence. With regard to planetesimal formation, dust concentrations in the sub-rings may remain smaller than the threshold for gravitational collapse. However, gas photoevaporation enhances dust settling and (partially) radial concentration, eventually triggering the formation of dust clumps of increasingly large density, in both the viscous and inviscid cases.}
{We conclude that planetesimal formation within dust-trapping pressure bumps is favored in very low-viscosity disks at late evolutionary stages, when sufficient gas has been removed by photoevaporation. This result is consistent with the inferred late formation of the parent bodies of chondritic meteorites in the Solar System.}

\titlerunning{Dust dynamics in disk dust traps}

\authorrunning{Tatarelli, M., et al.} 

\maketitle

\section{Introduction}

The model currently favored for the formation of planetesimals in a protoplanetary disk is the streaming instability (\citealt{youdin2005streaming}; see \citealt{squire2020physical} for a review). This instability arises from the relative drift between dust particles (mm or cm in size) and gas with mutual aerodynamic coupling and concentrates dust into filaments during a linear growth phase, where dust clumps can then form via a nonlinear growth phase (\citealt{abod2019mass}; \citealt{li2021thresholds}). When the dust density in a clump reaches the Roche density, the clump gravitationally collapses under its own self-gravity, forming 10-100 km sized bodies, or planetesimals \citep{klahr2020turbulence}.

However, there is evidence that the radial drift of dust relative to gas is limited to a fraction of the disk’s lifetime, likely within the first Myr. Indeed, extrasolar protoplanetary disks remain rich in dust throughout their lifetime of $\sim$ 5 Myr, and their dust radius does not seem to shrink with time \citep{najita2018protoplanetary}, suggesting that dust radial drift has been blocked \citep{birnstiel2024dust}. Moreover, the Solar System presents two generations of planetesimals. The first generation, including the parent bodies of iron meteorites and other achondrites, did form in the first Myr \citep{spitzer2021nucleosynthetic}, but the parent bodies of chondritic meteorites formed much later, between 2 and 4 Myr after the birth of the protoplanetary disk (e.g., \citealt{neumann2024recurrent}). Despite this, there are pairs of achondrites-chondrites (e.g., Aubrites and enstatite chondrites, achondrites NWA 6704 and NWA 011 and CR chondrites, etc.) that have identical nucleosynthetic isotopic anomalies, which, given that the disk was radially heterogeneous (e.g. \citealt{kleine2020non}), implies that dust remained blocked and confined during the time lapse separating the two planetesimal formation events. Thus, the very framework of the streaming instability (dust radial drift) most likely does not apply to planetesimals forming in a late disk.

An efficient mechanism for trapping dust in the disk is a pressure bump. A local maximum in an otherwise monotonically decreasing radial pressure profile of the gas acts as a dust trap by halting the inward drift and allowing dust grains to accumulate. Protoplanetary disk observations by the Atacama Large Millimeter/submillimeter Array (ALMA) have revealed that concentric dust rings are a ubiquitous feature among these observed disks \citep[see, e.g.,][]{andrews2018disk} and are nearly all confined within pressure maxima \citep{stadler2025exoalma}. The question is then whether these dust rings can also act as the nursery for dust clumps and be the eventual birth place of planetesimals. To answer this question, we must consider the dust-gas dynamics within the dust rings and the hydro-instabilities that may be active and capable of concentrating dust into denser regions.

Although the dust is effectively trapped, the formation of these rings does not immediately trigger planetesimal formation. Turbulence in the gas prevents dust from settling sufficiently toward the midplane, a necessary condition for gravitational collapse. Even modest turbulence levels (with an $\alpha$-viscosity around $10^{-4}$) can inhibit efficient sedimentation. The same is true for the radial concentration at the pressure maximum \citep{dullemond2018disk}. Thus, a large amount of dust might need to accumulate over time into the ring, before planetesimal formation becomes possible. 

This highlights the importance of modeling dust rings with realistic levels of turbulence in order to fully understand how hydrodynamic instabilities behave within and near pressure bumps. While pressure bumps can effectively trap dust and suppress classical instabilities like the Kelvin-Helmholtz instability at their centers, the overall picture remains complex. The limited effectiveness of the streaming instability—particularly at the pressure maximum where relative dust-gas drift vanishes—suggests that additional mechanisms are needed to form dust clumps in these regions (see, e.g., \citealt{auffinger2018linear, carrera2022streaming}). This motivates the exploration of alternative instabilities that can operate under realistic disk conditions and potentially drive dust clumping within the dust rings. One such candidate is the Dusty Rossby Wave Instability (DRWI), first coined by \cite{liu2023dusty}, and investigated in \cite{Cui_2025}.

They identify two distinct modes of the DRWI. Type I is similar to the classic Rossby Wave Instability (RWI), but with dust. This mode leads to the formation of a single large gas vortex that traps and concentrates dust—a process that has been well studied. However, the viability of Type I DRWI faces several challenges in realistic disk conditions.

First, the classic RWI requires a relatively sharp pressure bump, whereas milder pressure bumps are expected to be more common in protoplanetary disks. Second, as previously mentioned, observations show that dust rings are a ubiquitous feature of these disks, while crescent-shaped asymmetries, which are thought to be signatures of gas vortices, are comparatively rare. The widespread presence of dust rings also suggests that they are long-lived structures. In contrast, as Type I DRWI leads to vortex formation, it disrupts and destroys the dust ring on timescales that are much shorter than the lifetime of dust rings inferred from observations.

This motivates a more in-depth examination of the newly identified Type II DRWI, which presents a promising alternative pathway to planetesimal formation. One key advantage is that it can operate in relatively mild pressure bumps—including many that remain stable to the standard RWI—making it a more broadly applicable and plausible scenario. Additionally, this instability has the potential to concentrate dust into dense clumps, which may eventually form planetesimals, while preserving the overall dust ring structure.

In this work, after presenting our computational methods (Section~\ref{sec:methods}) and disk model (Section~\ref{sec:disk_model}), we first reproduce the results of \cite{liu2023dusty} on the Type II DRWI and extend them to longer timescales (Section~\ref{sec:2D_global}). We then extend the analysis to 3D, beginning with a viscous disk with $\alpha=10^{-4}$ (Section~\ref{sec:3d_viscous}), followed by an inviscid disk (Section~\ref{sec:3d_inviscid}). In the inviscid case, we consider both a setup with a finite dust reservoir (no inflow) and a scenario with continuous dust inflow toward the pressure bump. The latter is investigated using a high-resolution 2D $r-z$ simulation, and we show that it is an inefficient mechanism for increasing the local dust density, particularly in the inviscid regime. Moreover, as discussed above, sustained dust inflow over the disk lifetime is unlikely. For these reasons, in Section \ref{sec:gas_evap} we study an alternative mechanism for planetesimal formation in a pressure bump, i.e., during gas removal. In particular, we investigate how the maximal dust density should scale with the overall dust-to-gas mass ratio in the ring, which should eventually lead to a gravitational instability of the dust clumps as gas is removed. The conclusions will follow in Section \ref{sec:conclusions}.

\section{Numerical methods} \label{sec:methods}
We consider a global protoplanetary disk model  treated as a non-self-gravitating mixture of gas and dust whose motion is described by the Navier-Stokes equations. 
We use spherical coordinates $(r,\phi,\theta)$ 
where $r$ is the radial distance from the host star
of mass $M_*$  (which is at the origin of the coordinate system), $\phi$ the 
polar angle measured from the $z$-axis (the colatitude), and $\theta$ is the azimuthal coordinate measured from the $x$-axis. The midplane of the disk is located at the equator
$\phi = \frac {\pi} {2}$. The gravitational potential of the central star acting on the disk is $\Phi=-GM_{*}/r$. We do not consider indirect terms that arise from the primary acceleration due to disk's gravity \citep{Cridaetal2025}.\par
We use the code fargOCA \citep{Legaetal2014}  which is a grid-based code that solves the hydrodynamic equations with a time-explicit method, using operator splitting and upwind techniques. The code has been recently reorganized to run both on CPUs and GPUs  with the addition of dust dynamics  (Miniussi et al. in prep.).

 We model the gas as a fluid  with volume density 
 $\rho_g$ 
 and velocity $\vec u=(u_r,u_{\theta},u_{\phi})$  with $u_{\theta}=r\sin (\phi)(\omega+\Omega_f)$ 
where $\omega$ is the azimuthal angular velocity in a frame  rotating around the z axis at angular velocity $\Omega_
f$. The dust is modeled as a non-viscous pressureless fluid described by a volume density $\rho_d$ and a velocity $\vec {\rm v}$. Dust experiences a drag force from the collisions with gas that can be expressed (per unit volume) as: $$\vec F_d=-\rho_d \frac{ \Omega_k}{S_t}(\vec {\rm v}-\vec u)$$ where  $\Omega_k/S_t$ is the friction timescale  which characterizes the time needed for  dust grains with Stokes number $S_t$ to adjust their velocity to a change of gas velocity. The gas feels a reaction force, or feedback, of sign opposite to drag force exerted on the dust which is proportional to the amount of dust: $$\vec F_g=\rho_d \frac{ \Omega_k}{S_t}(\vec {\rm v} - \vec u)$$ 
The continuity  equation for the gas reads: $$\frac {\partial \rho_g} {\partial t}+ \nabla \cdot (\rho_g \vec u)= 0 $$ 
The Navier-Stokes equations  for  the radial momentum  $Y_r= \rho_g u_r$, the polar momentum $Y_\phi = \rho_g r u_\phi$ and the angular momentum $Y_\theta = \rho_g r\sin(\phi) u_\theta$ read:
\begin{equation}\label{ns_gas}
\left\lbrace \begin{array}{lll}
\frac{\partial Y_r } {\partial t}+ \nabla  \cdot (Y_r\vec u) & = & \rho_g [\frac {u_\phi 
^2} { r }+ \frac{u_\theta^2}{r} -{\frac {\partial \Phi} { \partial r}} +{\frac {1} {\rho_g}} (f_r-{\frac {\partial P} {\partial r}}) \\
 &   &  +\frac{\rho_d}{\rho_g}\frac{ \Omega_k}{S_t}(v_r-u_r)] \\
{\frac {\partial Y_\phi} {\partial t}}+ \nabla \cdot (Y_\phi \vec u) & = &
 \rho_g r[\frac {u_\theta^2\cot(\phi)}{r} -\frac {1}{r\rho_g}  
 ( f_\phi -{\frac {1}{r} }{\frac {\partial P} {\partial \phi}} )  \\ & & + \frac{\rho_d}{\rho_g}\frac{ \Omega_k}{S_t}(v_{\phi}-u_{\phi})] \\
{\frac {\partial Y_\theta} {\partial t}}+ \nabla \cdot (Y_\theta \vec u) & = &
 \rho_g r \sin (\phi)[\frac {1}{\rho_g}(f_\theta 
 -{\frac {1}{r \sin \phi}}{\frac {\partial P}{\partial \theta}})+F_{\theta}  \\& & +\frac{\rho_d}{\rho_g}\frac{ \Omega_k}
{S_t}(v_{\theta}-u_{\theta})]
\end{array} \right.
\end{equation}
The function $f=(f_r,f_\phi,f_\theta)$ is the divergence of the
stress tensor (see, for example, \cite{Tassoul78}). The gravitational potential $\Phi$  depends on $r$ and therefore contributes to a radial acceleration in the above equations. The term $F_{\theta}$ is a forcing term that will be defined in Section \ref{sec:disk_model}. The fluid equations are closed by the definition of an equation of state (EoS). In this paper we consider a locally isothermal disk with pressure: $P=c^2_s \rho_g$ where the sound speed $c_s$ is given by $c_s=\Omega_k H_g$ with the pressure scale height $H_g=h_0r$ and $h_0$ the disk aspect ratio.
\par
The continuity equation for the dust reads:
$\frac {\partial \rho_{d} } { \partial t}  + \nabla\cdot (\rho_{d}{\vec {\rm v}}) = \nabla \cdot \vec j$
with
\begin{equation}\label{eq:dust_diffusion}
    \vec j = D_d \rho_g \nabla (\frac {\rho_d}{\rho_g})
\end{equation}
where $D_d$ is the dust  diffusion coefficient \citep{liu2023dusty}:
$D_d = \rho_g/(\rho_g+\rho_d) D_{g}$ with $D_{g}=\alpha H_g^2 \Omega_k$ and $\alpha$ is the Shakura–Sunyaev turbulent viscosity parameter related to the disk's viscosity $\nu$ by $\nu = \alpha c_s H$. This formula for the dust diffusion coefficient can be simply derived by assuming that for a turbulent fluid, the turbulent kinetic energy is $1/2 \rho_g u_{turb}^2$. But if we add $\rho_d$, the energy becomes $1/2 (\rho_g+\rho_d) {u}'^2_{turb}$ where ${u}'_{turb}$ is the new turbulent velocity. For the energy to be the same (the only source of energy comes from the turbulence of the gas, the dust is just an inertia), one has $${u}'_{turb}=\sqrt{ \rho_g/(\rho_g+\rho_d)}{u} _{turb}.$$ In other words, the turbulent velocity is quenched by a factor $\sqrt{\rho_g/(\rho_g+\rho_d)}$. We notice that the diffusion coefficient is related to the turbulent velocity ${u}_{turb}$ by the formula $D_g=u_{turb}^2/\Omega_k$ so that, the dust diffusion coefficient 
is quenched by a factor $\rho_g/(\rho_g+\rho_d)$.  \par
The Navier-Stokes equations  for the dust's radial momentum  $J_r= \rho_d {\rm v_r}$, the polar momentum $J_\phi = \rho_d r {\rm v}_{\phi}$ and the angular momentum 
 $J_\theta = \rho_d r\sin(\phi) {\rm v}_{\theta}$  read:

\begin{equation}
  \label{ns_dust}
\left\lbrace \begin{array}{lll}
{\partial J_r \over \partial t}+ \nabla  \cdot (J_r\vec {\rm v}) & = & \rho_d [{{\rm v}_\phi 
^2 \over r }+ {{\rm v}_\theta^2\over r} -{\partial \Phi \over \partial r} -\frac{ \Omega_K}{\St}({\rm v}_r-u_r)] \\
{\partial J_\phi \over \partial t}+ \nabla \cdot (J_\phi \vec {\rm v}) & = &
 \rho_d r[{{\rm v}_\theta^2\cot(\phi)\over r}   - \frac{ \Omega_K}{\St}({\rm v}_\phi-u_\phi)] \\
{\partial J_\theta \over \partial t}+ \nabla \cdot (J_\theta \vec {\rm v}) & = &
-\rho_d r \sin (\phi)[\frac{ \Omega_K}{\St}({\rm v}_\theta-u_\theta)]
\end{array} \right.
\end{equation}

\section{Disk model}
\label{sec:disk_model}
We consider  disks 
with background density  profile  
\begin{equation} \label{eq:sigma_gas_background}
    \Sigma_b (r) = \Sigma_0 (r/r_0)^{-p},
\end{equation}
where $\Sigma_0$ is the gas surface density at $r/r_0=1$ and $p$ is the power law slope while $r_0$ is the unit of distance in au. In all of our simulations, we apply the following parameters: $\Sigma_0=10^{-4}$, $p=1$ and aspect ratio: $h_0=0.05$. The code units are such that $G=M_*\equiv M_\odot=1$, and the unit of distance $r_0$ is arbitrary in au.
The unit of time is  $(r_0/{\rm au})^{3/2}/(2\pi)\,{\rm yr}$. 
For our model we adopt as the unit of length $r_0=1$~au. We call $P_{\rm{orb}}$ the orbital period at $r=r_0$. For simplicity in the following we will omit the $r_0$ term. In the 3D model we define the background volume density from hydrostatic equilibrium in cylindrical coordinates $(R,z)$ as:
\begin{equation}\label{eq:rho_gas_background}
    \rho_b(R,z) = \frac{\Sigma_0}{h_0 \sqrt{2\pi}}R^{-p-1}\exp-\left(\frac {z}{2H_g}\right)^2
\end{equation}
where $z=r\cos(\phi)$ and $R=r\sin(\phi)$ with $R \sim r$ in the thin disk approximation used in this paper. 

\subsection{Density bump - initialization and forcing term}

Following a similar approach to \cite{liu2023dusty}, we model the gas density bump as a Gaussian profile and maintain it by applying a torque via a forcing term $F_{\theta}$ in the azimuthal direction that balances viscous diffusion.

We introduce the density bump in an otherwise monotonically decreasing radial gas density profile. We are mainly interested in the 3D model, however, we will start our discussion with a simulation in the 2D $(r,\theta)$ case and therefore we define both the density bump and the forcing term in the two cases. Precisely, in the case of a 2D vertically integrated disk, we set an initial surface density profile of the form:
\begin{equation}\label{eq:density_bump_2D}
    \Sigma_{g}(r) = \Sigma_b \exp\left[A\exp\left(-\frac {(r-r_{bump})^2}{2 \Delta w^2}\right)\right],
\end{equation}
where $A$ is the bump amplitude and  $\Delta w$ is the bump width. The position of the bump maximum is $r_{bump}$.
In 3D, the gas density profile takes the following form, with the same bump definition as in 2D (Eq. \ref{eq:density_bump_2D}):
\begin{equation}\label{eq:density_bump_3D}
        \rho_{g}(r,\phi) = \rho_b \exp\left[A\exp\left(-\frac {(r-r_{bump})^2}{2 \Delta w^2}\right)\right],
\end{equation}

To maintain the density bump against viscous diffusion, we have introduced a forcing term in the azimuthal component of Eq.\ref{ns_gas}. Since the pressure has no azimuthal dependence  to compensate the viscous diffusion the forcing term is:
\begin{equation}
    F_{\theta} = -\frac{1}{d}f_{\theta},
\end{equation}
with  the  function $d$ corresponding to the surface density $\Sigma_g$ or the  volume density $\rho_g$ according to
 the dimension of the problem and 
$f_{\theta}$ is  the azimuthal component of the divergence of the stress tensor.
In the 3D case $f_{\theta}$ is given by: 
\begin{multline}\label{eq:diverg_stress_tensor_azi}
    f_{\theta} = \frac{1}{r \sin{\phi}} \left[ \frac{\sin{\phi}}{r} \frac{\partial}{\partial r}(r^2 \tau_{\theta r})
 + \frac{\partial}{\partial \phi}(\tau_{\phi \theta} \sin{\phi}) + \frac{\partial \tau_{\theta \theta}}{\partial \theta} \right] \\ 
    + \frac{\tau_{r \theta}}{r}
    + \frac{\tau_{\theta \phi} \cot{\phi}}{r},
    \end{multline}
with the components of the viscous stress tensor from \cite{Tassoul78} simplified with the condition $u_r=0$ and $u_{\phi}=0$:
\begin{equation}\label{stresstensor}
\left\lbrace \begin{array}{lllll}
    \tau_{\theta r} & = &\tau_{r \theta} & = &\rho_g \nu r \frac{\partial}{\partial r}(\frac{u_{\theta}}{r}) \\
    \tau_{\phi \theta} & =&  \tau_{\theta \phi} &=& \rho_g \nu \frac{\sin{\phi}}{r} \frac{\partial}{\partial \phi}(\frac{u_{\theta}}{\sin{\phi}}) \\
    \tau_{\theta \theta} & & & = &  2\rho_g\nu ({1\over  r\sin \phi}{\partial u_\theta \over \partial \theta})
\end{array} \right.
\end{equation}

Considering that $u_{\theta}$ has no azimuthal dependence, we have $\tau_{\theta \theta}=0$, and the forcing term finally reads:

\begin{multline} \label{eq:bump_forcing_3D}
    F_{\theta} = - \frac{1}{\rho_g} \Bigg[ \frac{1}{r^2}\frac{\partial}{\partial r}(r^2 \tau_{\theta r}) + \frac{\tau_{r \theta}}{r} \\
    + \frac{1}{r \sin{\phi}} \frac{\partial}{\partial \phi}(\tau_{\phi \theta} \sin{\phi}) + \frac{\tau_{\theta \phi} \cot{\phi}}{r} \Bigg].
\end{multline}
By neglecting the terms depending on the colatitude in Eq.\ref{eq:bump_forcing_3D} we have  the forcing term in the 2D case:
\begin{equation} \label{eq:bump_forcing}
    F_{\theta} = - \frac{1}{\Sigma_g} \left( \frac{1}{r^2}\frac{\partial}{\partial r}(r^2 \tau_{\theta r}) + \frac{\tau_{r \theta}}{r} \right),
\end{equation}
where $\tau_{\theta r} = \tau_{r \theta} = \Sigma_g \nu r \frac{\partial}{\partial r}(u_{\theta}/r)$.\\

We build an equilibrium disk by considering the initial radial velocity $u_r=0$ and the azimuthal velocity from the centrifugal balance:
\begin{equation}
    u_{\theta}^2/r = \Omega^2_k r + \frac{1}{d} \frac {\partial P} {\partial r}
 \end{equation}
 and $P=c_s^2d$. In the 3D case we set the initial polar velocity to zero. 

\subsection{Initial conditions for dust}

We recall that the gas rotates at a slightly sub-Keplerian speed and the relative difference with respect to the Keplerian speed ${\rm v}_{k}$ is indicated by  $\eta \equiv ({\rm v}_k-u_{\theta})/{\rm v}_k$, which can be computed as:
\begin{equation}
\label{eta}
\eta = -\frac{1}{2}\left(\frac{H}{r}\right)^2 \frac {\partial \log P}{\partial \log r},
\end{equation}
with $\eta \propto h_0^2$ in usual power law disks. The radial dust velocity component $v_r$ is then initialized as \citep{Takeuchi2002}:
\begin{equation}
\label{eq:vrdust}
    {\rm v}_r = \frac{u_r}{S_t^2 + 1} - \frac{2S_t} {S_t^2 + 1}\eta {\rm v}_K.
\end{equation}
We remark that the gas velocity  $u_r$ is expected to be small. Neglecting $u_r$  if the pressure gradient is negative then  $\eta >0$ and  the dust moves toward the star, while in the case of a  positive pressure gradient  ($\eta <
0$) the situation is reversed.  In the presence of a pressure maximum ($\eta =0$), the dust can be trapped. The initial azimuthal velocity is:
\begin{equation}
    {\rm v}_\theta = u_\theta - \frac{1}{2} S_t {\rm v}_r,
\end{equation}
while the initial polar velocity is zero as for the gas. The dust density is initialized as a fraction  $\epsilon_0$ typically of a few percent of the gas density.

\subsection{Boundary conditions}\label{sec:boundary}

We provide the values of dust and gas components in the ghost cells (for the radial and vertical directions) according to values of the closest neighbor or first active cell.  We consider a global disk with periodic azimuthal boundary conditions. For the gas components in the radial direction we use the classical prescription \citep{deValBorroetal2006} of evanescent boundary condition. \par

For the dust component, we implement an open inner radial boundary, allowing dust to drift out of the computational domain. Specifically, dust with negative radial velocity in the first active cells is advected into the ghost cells. To prevent dust pile-up, we copy the density from the first active cells into the ghost cells. The azimuthal velocity component of the first active cells is extrapolated using a Keplerian profile within the ghost cells.
At the outer radial boundary, the azimuthal velocity of the last active cells is similarly extrapolated using a Keplerian profile in the ghost cells. For the dust density and radial velocity, we consider two distinct boundary conditions:

\begin{itemize}
\item {\it no inflow } : The outer part of the simulation domain depletes according to radial dust drift towards the pressure bump. This set-up is intended to mimic the case where a barrier in the disk beyond the simulation domain halts dust flow toward the considered pressure bump so that only the local dust can concentrate there. Here, the radial velocity in the ghost cells is set to zero.

\item {\it inflow}: We impose a constant dust inflow by setting the ghost cell’s radial velocity to the theoretical value (Eq. \ref{eq:vrdust}) and resetting its density to the initial profile. This simulates a continuous replenishment of the dust reservoir in the simulation domain via advection and diffusion. In 3D simulations, we additionally account for dust settling by renormalizing the vertical density distribution in the ghost cells at each time step, using the vertical profile of the first active cell.
\end{itemize}

\par At the vertical boundaries, we enforce a no-flow condition by setting the vertical velocity component to zero for both gas and dust, preventing any inflow or outflow from the domain. All other quantities in the ghost cells are directly copied from the nearest active cells.

\section{2D simulation results}\label{sec:2D_global}

\begin{figure*}[htbp]
    \centering
    \includegraphics[width=\textwidth]{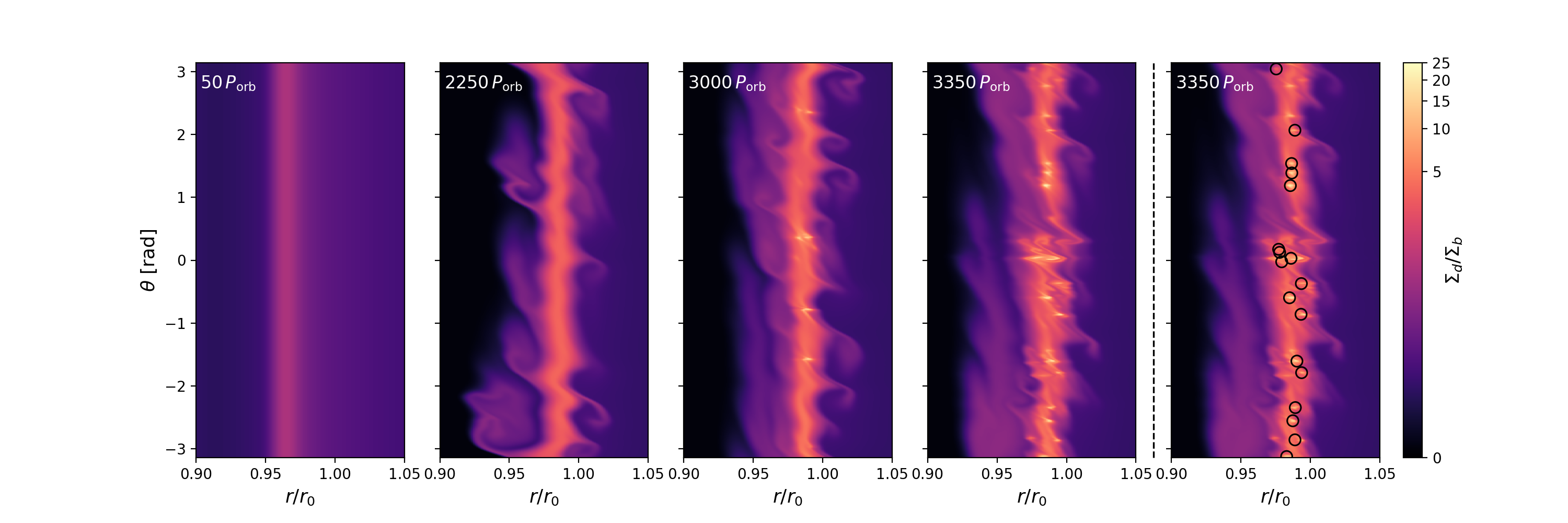}
    \caption{Snapshots in the 2D Type II DRWI run of the dust density in the azimuthal-radial plane. The time is annotated on the top left of each panel. The panels are colored in a power-law scale, and the dust surface density is plotted with respect to the initial background gas density profile (Eq. \ref{eq:sigma_gas_background}). The panel at 3350 $P_{\rm orb}$ is repeated to the right of the dotted black line, with the identified dust clumps circled in black.}
    \label{fig:2D_Type2_density_snapshots}
\end{figure*}

The first goal of this paper is to reproduce the 2D local shearing-box results of \cite{liu2023dusty}, hereafter LB23, using 2D global disk simulations.  We will begin by describing the initial disk setup and then present our findings for Type II DRWI, highlighting similarities and differences with LB23.

 We begin from an equilibrium disk with a homogeneous global dust-to-gas density ratio $\epsilon_0=0.02$ and viscosity $\alpha=10^{-4}$, and initialize a gas density bump centered at $r_{bump}\equiv r_0 = 1$, with amplitude $A=0.8$ and width $\Delta w/h_0=1.5$ (the same density bump parameters and viscosity used in LB23's fiducial Type II run), which is expected to be stable to the classic RWI \citep{chang2023origin}. The radial simulation domain spans the interval $[0.6,1.4]$ and we consider $(N_r \times N_\phi)=(1024\times3072)$ grid cells.

\begin{figure}[htbp]
    \centering
    \includegraphics[width=0.42\textwidth]{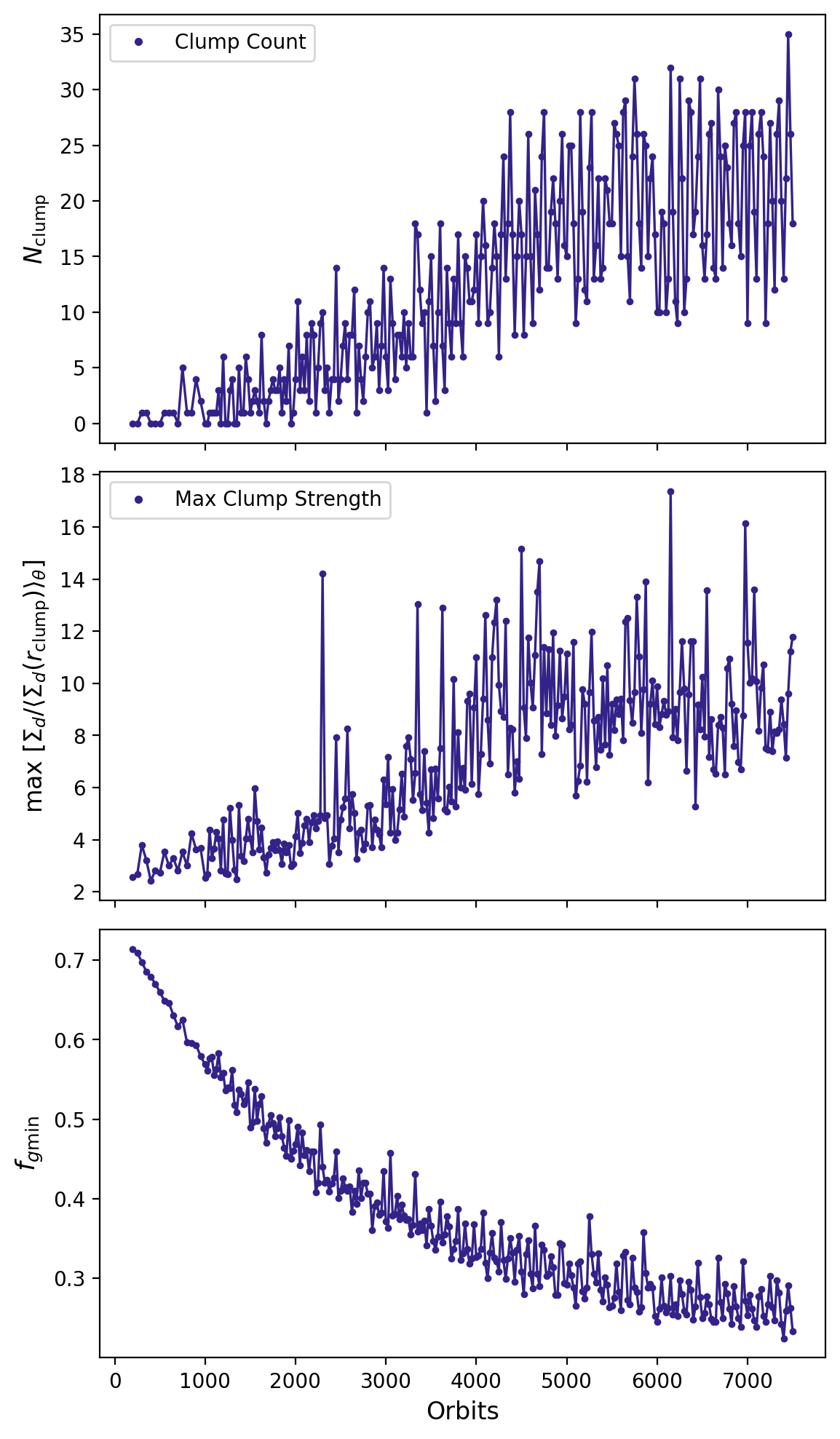}
    \caption{Evolution of the dust clumps in the 2D Type II DRWI run. Top panel: Number of identified dust clumps at each time; Middle panel: Maximum density of the dust clumps at each time, normalized by the average dust density at the clump's radial location; Bottom panel: Evolution of the minimum gas-to-dust ratio.}
    \label{fig:dust_clump_trends}
\end{figure}

The established pressure bump reverses the sign of $\eta$ (Eq. \ref{eta}) and traps the inward drifting dust and forms the dust ring, locally increasing the dust-to-gas ratio within the ring. The initial condition of our disk differs from that of LB23, as they begin from a pre-established dust ring in equilibrium which has sufficient dust-to-gas ratio to trigger Type II DRWI, once perturbed. We instead impose dust \textit{inflow} from the outer radial boundary (according to Section \ref{sec:boundary}) to gradually increase the dust-to-gas ratio, ensuring that at the pressure bump we eventually reach the critical dust-to-gas ratio required for the dust ring to become unstable. Following LB23 we characterize the dust enrichment at the pressure bump by:
\begin{equation}
    f_{g\text{min}} = \min\left( \frac{\bar\Sigma_g}{\bar\Sigma_g+\bar\Sigma_d}\right),
\end{equation}
where $\bar\Sigma_g$ and $\bar\Sigma_d$ are azimuthally averaged quantities, and the minimum is taken over the simulation domain and occurs near the pressure maximum. We find that the Type II DRWI is triggered once $f_{g\text{min}}$ decreases below 0.5 (see Fig. \ref{fig:dust_clump_trends}). LB23 found a threshold at $f_{g\text{min}}= 0.536$. Given the differences in the simulation set-up our results are satisfactorily similar.

Figure \ref{fig:2D_Type2_density_snapshots} shows the evolution of the dust ring after the onset of the Type II DRWI. After the ring becomes unstable, dust clumps eventually form, which can be seen in the panels at 3000 and 3350 $P_{\rm orb}$. The panel at 3350 $P_{\rm orb}$ is repeated with the dust clumps circled in black. We identify the dust clumps by applying an algorithm to find the local dust density maxima; then the maxima exceeding three times the azimuthal average are considered to be dust clumps. Although LB23 note the transient nature of the dust clumps in their simulations, we extend their analysis by examining the behavior and evolution of these clumps when dust is continuously supplied to the ring. Figure \ref{fig:dust_clump_trends} presents the evolution of the number of dust clumps in the ring (top panel), the maximum dust clump strength (middle panel), and the evolution of $f_{g\text{min}}$. We define the clump strength as max[$\Sigma_d / \bar\Sigma_d(r_{\rm clump})$], with $\bar\Sigma_d(r_{\rm clump})$ being the azimuthal average of the dust density at the clump's location. We find that both the number of dust clumps and the maximum density of the clumps increase as more dust is collected in the ring (i.e., as $f_{g\text{min}}$ decreases).

The instability reaches saturation near $4500$ $ P_{\rm orb}$, after which the number and strength of the dust clumps continue to fluctuate, but show no significant increase. The number of dust clumps averages at 20-25 (Fig. \ref{fig:dust_clump_trends} top panel) and the dust clump strength around 10 (Fig. \ref{fig:dust_clump_trends} middle panel), corresponding to $\Sigma_d/\Sigma_b\approx30$, where $\Sigma_b$ is defined in Eq. \ref{eq:sigma_gas_background}. The maximum dust clump strength we find is close to 18, which corresponds to $\Sigma_d/\Sigma_b\approx34$.

Although our 2D simulations produce dust clumps, assessing whether they reach the Hill density, the threshold for gravitational collapse, is difficult for two reasons: first, a fully 3D treatment is required to measure the volume density of dust on the midplane (although some estimates can be used); second, because our code does not include self-gravity, all resulting densities scale with the value of $\Sigma_b$ we assume. Nevertheless, it is important at this stage to extend the analysis to 3D simulations, as different behaviors and instabilities may appear.

\section{3D simulation results - viscous disk}\label{sec:3d_viscous}

\begin{table*}[t]
\centering
\caption{Summary of 3D simulation parameters.}
\begin{tabular}{lcccccc}
\hline\hline
Run & Dust Inflow (BC) & $(N_r \times N_\phi \times N_\theta)$ & $\alpha$ & $\St$ & $\langle \epsilon_0 \rangle$ \\
\hline
Viscous, stratified & \checkmark & $393\times60\times3072$ & $10^{-4}$ & 0.03 & 0.05 \\
Viscous, unstratified & \checkmark & $393\times60\times3072$ & $10^{-4}$ & 0.03 & 0.05 \\
Inviscid (\textit{no inflow}) & $\times$ & $393\times60\times3072$ & 0 & 0.03 & 0.05 \\
Inviscid (2D $r-z$, \textit{no inflow}) & $\times$ & $393\times60$ & 0 & 0.03 & 0.05 \\
Inviscid (2D $r-z$, \textit{inflow}) & \checkmark & $3930\times600$ & 0 & 0.03 & 0.01 \\
Viscous, gas evap & $\times$ & $393\times60\times3072$ & $10^{-4}$ & 0.03 & -- \\
Viscous, gas evap (2D $r-z$) & $\times$ & $3930\times600$ & $10^{-4}$ & 0.03 & -- \\
Inviscid, gas evap & $\times$ & $393\times60\times3072$ & 0 & 0.03 & -- \\
\hline
\end{tabular}
\tablefoot{$A=0.8$, $\Delta w/h_0=1.5$, $r_{\text{bump}}=1$, $\Sigma_0=10^{-4}$, $p=1$, $h_0=0.05$.
All simulations use the same initial gas and dust surface density. The radial domain spans $r \in [0.8, 1.2]$, the colatitude spans $\pm 0.6h_0$ about the midplane, and the disk spans the full azimuthal range $\theta \in [0, 2\pi]$ in all cases except for the 2D $R$–$Z$ simulations. The second column indicates whether there is dust inflow from the outer radial boundary.
}
\label{tab:sim_params}
\end{table*}

Here, we expand on the work of LB23 by performing 3D global disk simulations. The vertical domain spans the $\phi$ interval $[1.54,1.60]$ ($\phi=\pi/2$ being the midplane), which is $\pm 0.6h_0$ about the midplane (much larger than the scale height of the dust layer), and the radial domain extends over $[0.8,1.2]$. The full $2\pi$ range in $\theta$ is considered. To begin, the setup is identical to the 2D case, but with an initial homogeneous global dust distribution of $ \epsilon_0 = 0.05$. Since 3D simulations are computationally expensive, we have used this unusually large $\epsilon_0$ value to reach the critical $f_{g{\rm min}}$ for DRWI Type II in fewer orbital periods than in 2D. The dust is then allowed to settle self-consistently and drift radially under the hydrodynamics of the code. Again, we impose dust \textit{inflow} from the outer radial boundary, as described in Section \ref{sec:boundary}. The gas density bump in 3D is established according to Eq. \ref{eq:density_bump_3D} and enforced according to Eq. \ref{eq:bump_forcing_3D}.

To thoroughly study DRWI and the evolution of the dust ring in 3D, we perform a series of tests, first examining the case of a viscous disk for a comparison with the previous 2D results, and then considering a disk with vanishing viscosity.

\subsection{Stratified viscous disk}\label{sec:3D_viscous_stratified}

\begin{figure}[htbp]
    \centering
    \includegraphics[width=0.48\textwidth]{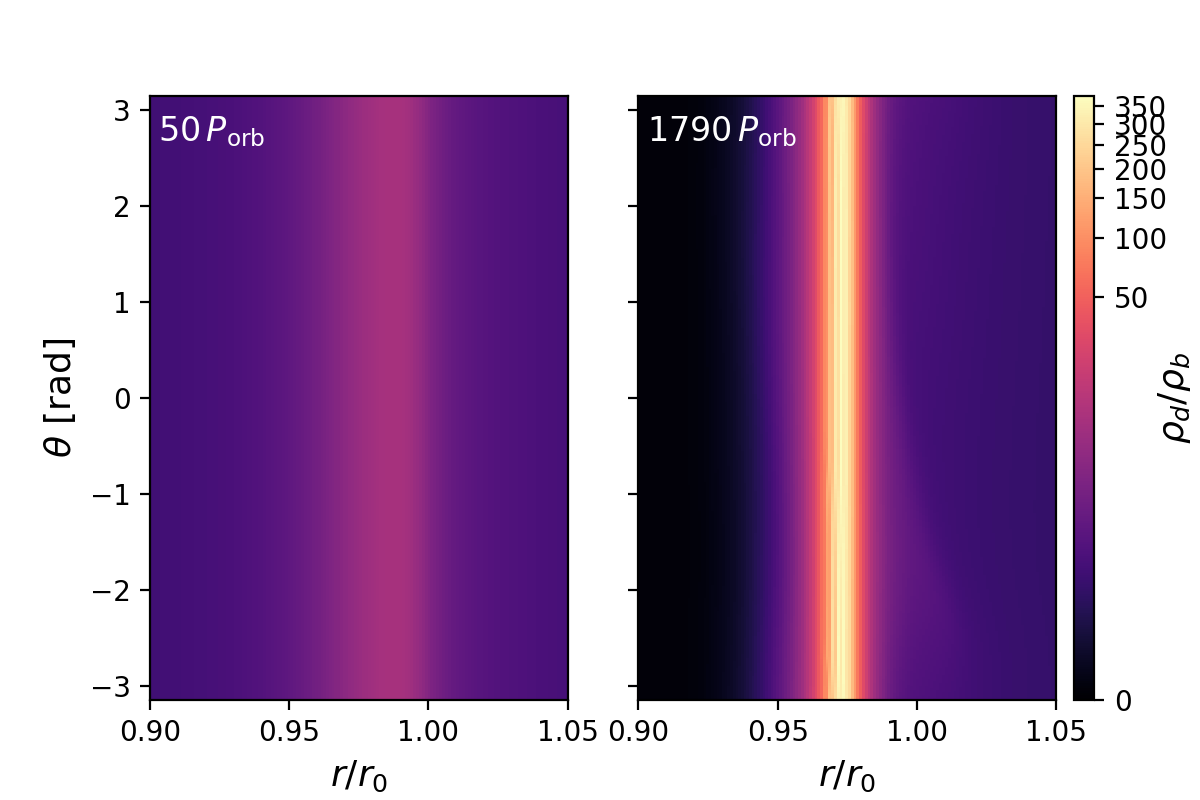}
    \caption{Similar to Fig. \ref{fig:2D_Type2_density_snapshots}, but depicting the volume density of dust on the midplane in a 3D simulation with the same pressure bump parameters and gas viscosity as in the 2D model of Section \ref{sec:2D_global}.}
    \label{fig:3D_viscous_density_snapshots}
\end{figure}

Our first test is with a 3D vertically stratified viscous disk. Figure \ref{fig:3D_viscous_density_snapshots} shows the dust density after 1790 $P_{\rm orb}$. No instability or dust clumping develops, even though $f_{g{\rm min}}$ decreases well below the critical value required to trigger Type II in 2D, and we find the same result in tests with a larger Stokes number $\St = 0.1$. This suppression is likely a consequence of the full 3D structure: dust settling concentrates dust into a thin midplane layer, leaving the upper regions dust-poor and unable to develop or sustain the instability. Viscous coupling between vertical layers then allows the unperturbed gas above and below the midplane to damp and smooth out any perturbations that would otherwise develop in the dust-rich midplane, preventing the growth of the instability throughout the column. We test this interpretation in the following section using a 3D vertically unstratified test.

\subsection{Unstratified viscous disk}\label{sec:3D_viscous_unstratified}

To simulate an unstratified disk, we remove the vertical component of gravity on both dust and gas, and thus the gas is initialized with a homogeneous vertical density profile and the dust-to-gas ratio is the same at every $z$, so that each layer is identical. Dust settling does not occur because of the suppression of the vertical component of the gravitational force.

\begin{figure}[htbp]
    \centering
    \includegraphics[width=0.5\textwidth]{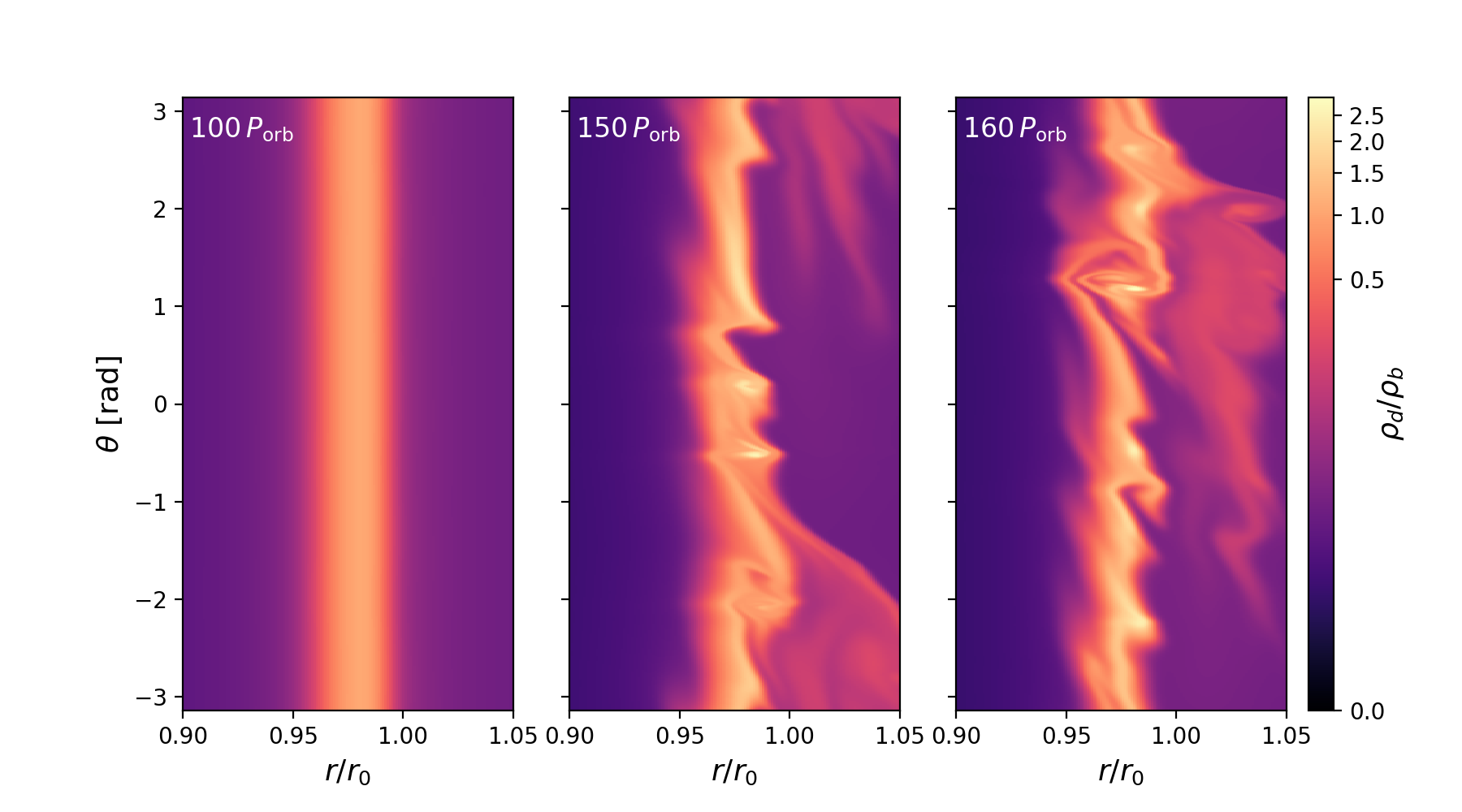}
    \caption{Same as Fig. \ref{fig:3D_viscous_density_snapshots}, but for the 3D unstratified simulation.}
    \label{fig:3D_viscous_unstrat_density_snapshots}
\end{figure}

Figure~\ref{fig:3D_viscous_unstrat_density_snapshots} shows that the dust ring in the unstratified case exhibits stronger perturbations than in the stratified run, even developing dust clumps and closely resembling the behavior seen in the 2D configuration (Figure \ref{fig:2D_Type2_density_snapshots}). Because dust is present throughout the vertical column, all layers become unstable, allowing perturbations to grow and dust clumps to form. This confirms our interpretation that, in the stratified disk, as dust settling concentrates solids in the midplane, it leaves the upper layers dust-poor and stable; viscous coupling between these layers then damps perturbations arising at the midplane, ultimately suppressing the instability. 

\section{3D simulation results - inviscid disk}\label{sec:3d_inviscid}

\begin{figure*}[htbp]
    \centering
    \includegraphics[width=\textwidth]{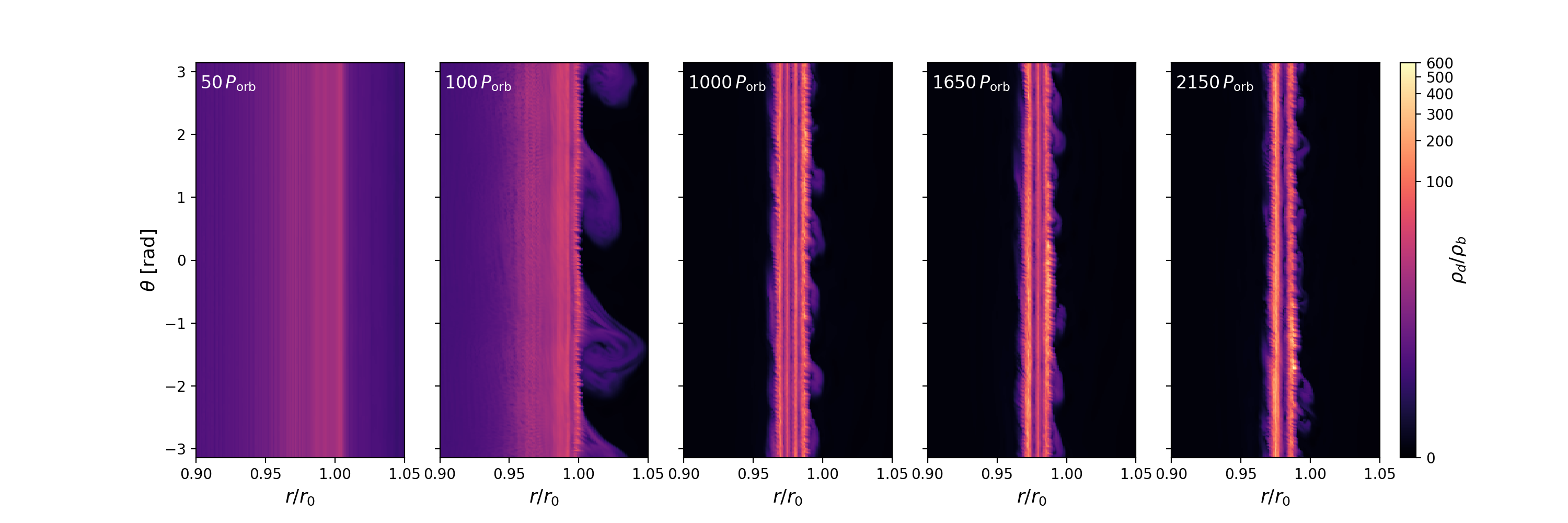}
    \caption{Similar to Fig. \ref{fig:3D_viscous_density_snapshots}, but for the 3D inviscid ($\alpha=0$) simulation without dust inflow.}
    \label{fig:3D_inviscid_noInflow_density_snapshots}
\end{figure*}

Given that the Type II DRWI is suppressed in 3D in a disk with $\alpha=10^{-4}$, we test what happens in the limit of an inviscid disk. For this purpose, we set $\alpha=0$. Of course, as all numerical codes, the simulation is still affected by some numerical viscosity, but various tests suggest that it is weaker than $\alpha=10^{-5}$ \citep{2021A&A...646A.166L}. Importantly, in the inviscid case, no diffusion is prescribed on the dust evolution. The only diffusion that is eventually present is that induced self-consistently by the gas dynamics. Furthermore, the function $f_\theta$ (Eq. \ref{eq:diverg_stress_tensor_azi}) is now identically null.

To investigate the role of dust supply, we consider two inviscid simulations that differ in their dust reservoir. In the first case, the disk contains a finite amount of dust with no inflow from the outer boundary, representing a situation in which dust has already drifted inward and accumulated in the pressure bump. Accordingly, we adopt a relatively high initial dust-to-gas ratio of $\epsilon_0=0.05$. In the second case, we allow for continuous dust inflow from the outer disk, starting from a standard interstellar value of $\epsilon_0=0.01$.

\subsection{No dust inflow}\label{sec:inviscid_noInflow}

We begin with a setup containing a finite amount of dust (\textit{no inflow}). In the absence of turbulent diffusion driven by $\alpha$-viscosity, dust is expected to settle into a thin, high-density midplane layer and to form correspondingly a narrower, sharper ring than in the viscous case. In this scenario, dust settling and radial contraction alone may be sufficient to reach the critical dust-to-gas ratio at the midplane required for instability within the ring.

Figure~\ref{fig:3D_inviscid_noInflow_density_snapshots} shows the evolution of the dust ring. Without viscous diffusion, multiple dust sub-rings emerge and interact dynamically. These sub-rings gradually merge, decreasing in number while concentrating the dust into a few dominant structures. Ultimately, two long-lived rings remain, spanning a finite radial width of $w_d = 0.02$. Moreover, neither of these rings become razor thin.

\begin{figure}[htbp]
    \centering
    \includegraphics[width=0.45\textwidth]{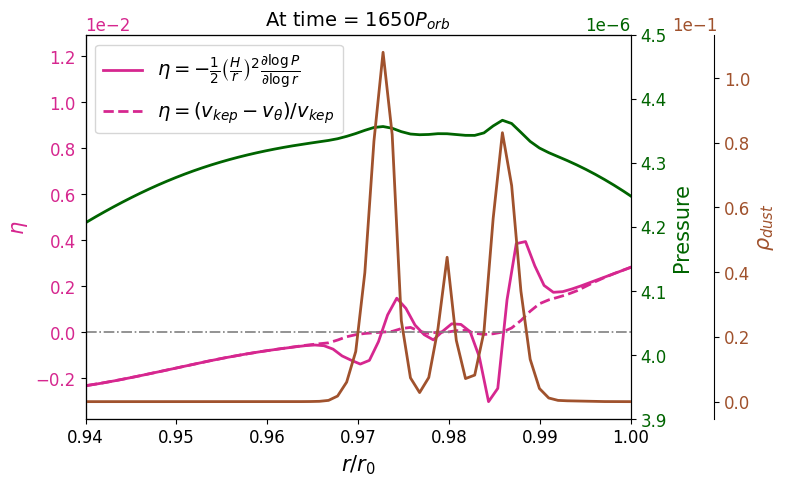}
    \caption{Radial profiles for the 3D inviscid simulation without dust inflow of the pressure (green), dust density (brown), and $\eta$ (pink). The solid pink curve is the pressure gradient, while the  dashed pink curve is the true value of $\eta$.}
    \label{fig:3D_inviscid_noinflow_eta}
\end{figure}

Figure~\ref{fig:3D_inviscid_noinflow_eta} illustrates the relationship between the dust-density peaks (brown), marking the locations of the rings, and the radial profile of $\eta$ (pink). The solid pink curve gives the pressure gradient, while the dashed pink curve provides the true value of $\eta$, defined as $\eta = (v_{\rm kep} - v_\theta)/v_\theta$. The latter is simply the former quenched by the factor $\rho_d/\rho_g$, as expected, and thus they have the same zero points. The zeros of $\eta$ (indicated by the horizontal dashed grey line) identify sign reversals, and those with positive slope coincide with the positions of the dust rings. In this high dust-to-gas ratio regime, the strengthened dust back-reaction, amplified by radial dust pile-ups and the radial dependence of the drift velocity, modifies the gas density profile and thus the pressure structure. Although the gas is only weakly compressible, this is sufficient, particularly when $\eta$ is small, to generate new zeros in the $\eta$ profile, where additional dust rings form. This happens when $\rho_d/\rho_g$ approaches 10. We discuss this in more detail in Section \ref{sec:inviscid_wInflow}.

We also checked that the inversions in the sign of $\eta$ occur also in a simulation with an adiabatic equation of state (instead of isothermal), with a cooling timescale of 100 orbital periods, consistent with the disk at 1 au. This is because the little compression needed to reverse the sign of $\eta$ produces negligible heating in the gas. Finally, we performed additional simulations with intermediate values of $\alpha=10^{-5}$ and $10^{-6}$. We find that dust sub-rings associated with $\eta$ sign inversions develop at $\alpha=10^{-6}$, whereas they are suppressed at $\alpha=10^{-5}$, with the evolution resembling that of the viscous case. This suggests that the transition occurs for $\alpha$ between $10^{-6}$ and $10^{-5}$.

\subsection{No dust inflow - 2D $r-z$ validation case}

Before proceeding with the next set of simulations, we perform a validation test to verify that we recover the same results (i.e. formation of multiple axis-symmetric rings) in a 2D $r-z$ simulation. 2D $r-z$ simulations are typical of most studies of the streaming instability (e.g. \citealt{li2021thresholds}) and offer the great advantage of enhancing resolution within the same computation time with respect to a full 3D simulation.

\begin{figure}[htbp]
\centering

\begin{subfigure}{0.49\columnwidth}
    \centering
    \includegraphics[width=\linewidth]{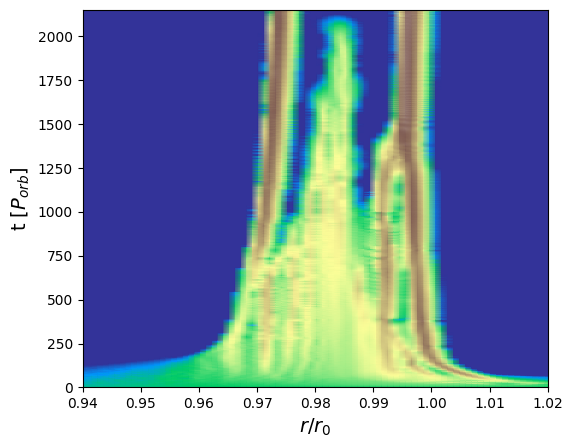}
    \caption{2D $r-z$}
\end{subfigure}
\hfill
\begin{subfigure}{0.473\columnwidth}
    \centering
    \includegraphics[width=\linewidth]{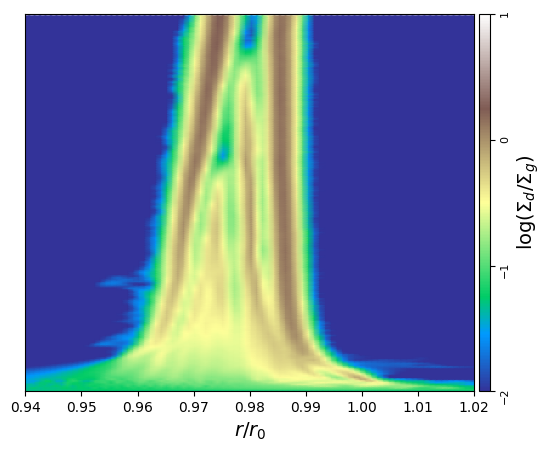}
    \caption{Full 3D}
\end{subfigure}

\caption{Comparison between the dust ring evolution in the 2D $r-z$ inviscid disk versus the full 3D disk with the same radial and vertical resolution.}
\label{fig:2Dvs3D_inviscid_noInflow}
\end{figure}

Figure~\ref{fig:2Dvs3D_inviscid_noInflow} shows the evolution of the dust surface density in a 2D $r-z$ disk (left) and in the fully 3D disk (right). Both simulations use the same set of parameters and resolution in the radial and vertical directions for a proper comparison, even though the 2D simulations can be run at much higher resolution. The results are qualitatively very similar, demonstrating that the 2D $r-z$ configuration captures the essential behavior observed in 3D. We therefore adopt the 2D setup for the following simulation in Section \ref{sec:inviscid_wInflow} in order to push the resolution as high as possible.

\subsection{Dust inflow}\label{sec:inviscid_wInflow}

\begin{figure*}[htbp]
    \centering
    \includegraphics[width=0.85\textwidth,height=18truecm]{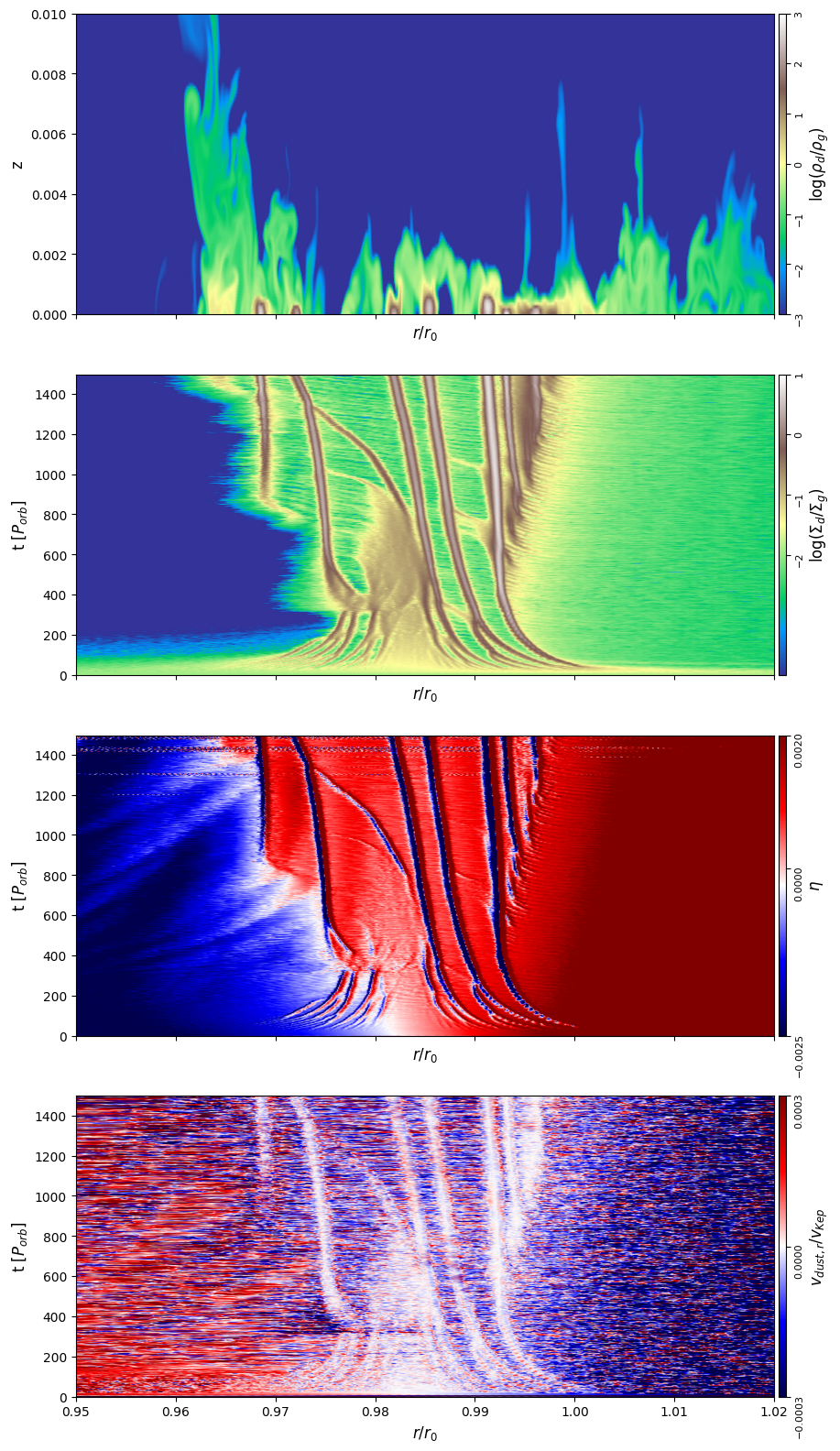}
    \caption{Evolution of the high resolution 2D $r-z$ inviscid disk with $\epsilon_0=0.01$ and dust inflow. In the top panel, we plot the $r-z$ view of the  dust to gas volume density at the final integration time $t=1500$ $P_{orb}$. The lower panels show the evolution with time versus $r/r_0$ of, respectively: the vertically integrated density ratio $\Sigma_d/ \Sigma_g$ (second panel), the $\eta$ values computed from midplane pressure gradient (third panel) and the midplane dust's radial velocity $v_r$ (bottom panel). In the third panel, a transition from blue to red indicates an $\eta$ sign inversion and thus a dust trapping location.}
    \label{fig:2DRZSigma}
\end{figure*}

Next, we examine the inviscid evolution under continuous dust inflow, starting from a more realistic initial dust-to-gas ratio of $\epsilon_0=0.01$. We adopt a 2D $r-z$ axisymmetric disk, implemented using a single grid cell in $\theta$, and otherwise use the same parameters as in the previous inviscid run (Section~\ref{sec:inviscid_noInflow}). The resolution is increased by a factor of 10 in both the radial and vertical directions. 
Figure \ref{fig:2DRZSigma} shows the resulting evolution with time versus the radial distance to the star of, respectively: the vertically integrated density ratio $\Sigma_d/ \Sigma_g$ (second panel), $\eta$ computed from the pressure gradient at the midplane (third panel) and the radial velocity of the dust at the midplane  $v_r$ (bottom panel). In the top panel of Fig. \ref{fig:2DRZSigma}, we show the vertical distribution of the dust at the end of the simulation, at $t=1500 P_{orb}$. In the second panel, we observe filaments of high density which correspond to the white regions along $z=0$ in the top panel.

\subsubsection{Multiple ring formation and streaming instability}\label{sec:SI}

Although the simulation starts from a lower dust-to-gas ratio, dust progressively drifts toward and accumulates at the pressure bump, and the resulting back-reaction becomes strong enough to reshape the gas profile. As in the \textit{no inflow} case, this generates multiple new locations where $\eta=0$ (Fig.~\ref{fig:2DRZSigma} third panel). The continuous supply of dust steadily modifies the gas density structure, partially flattening and shifting the original pressure bump. The partial flattening can be deduced from the second panel, which shows that the radial distribution of dust first contracts (in the first $\sim$ 300 orbital periods) then expands again. The shifting of the main pressure bump is shown in the third panel where the main boundary between the blue and red regions ($\eta$ < 0 and > 0 respectively) moves from $\sim$ 0.98 to $\sim$ 0.965 in the course of the simulation. This occurs because dust drift modifies the radial gas velocity through back-reaction, thereby reshaping the gas density and pressure profile. Remember that the azimuthal stress term $f_\theta$ (Eq.~\ref{eq:diverg_stress_tensor_azi}) vanishes because $\nu=0$.

As the pressure structure evolves, dust is able to drift inward of the original bump, transforming the ring into a non-uniform multi-ring configuration with localized regions of enhanced dust-to-gas ratio. New filaments continuously form and sometimes merge, each associated with new $\eta=0$ points. The dust radial drift velocity is essentially zero in the filaments (see the predominant white color in the bottom panel of Fig. \ref{fig:2DRZSigma}), meaning that dust is trapped in the filament and co-moving with it. However, because of turbulent fluctuations of the velocity (the red and blue dots in the vicinity of the main white filaments), some dust can diffuse out of the filament, and then starts to drift again until it is captured in a new filament.

The first two panels of Fig.~\ref{fig:2DRZSigma} show a pattern reminiscent of the streaming instability, which is expected as dust approaches the pressure maximum (initially located at $r=0.983$), slows down and accumulates on its outer side, progressively enhancing the local dust-to-gas ratio until the threshold for the streaming instability can be reached. The difference from the classic streaming instability is that, as soon as pronounced filaments appear, $\eta$ changes sign, as indicated by the appearance of blue stripes in the red domain of the third panel of Fig. \ref{fig:2DRZSigma}. The theory of the streaming instability is usually developed for an incompressible fluid, but in reality the gas is weakly compressed by the dust filaments, inducing oscillations of $\eta$ around its well defined positive value. In the vicinity of the pressure bump, $\eta$ is small, and the weak compression of the gas is enough to force it to change sign. Once this happens, the dust is trapped in each filament (see bottom panel, where the dust has zero radial velocity within the filaments). 

Finally, in the top panel of Fig. \ref{fig:2DRZSigma} we observe that the dust layer becomes increasingly vertically extended at radii $r>1$. In an inviscid disk, one would expect the dust to settle into an extremely thin midplane layer, yet this is not what the simulation produces. Although vertical stirring could in principle arise from the Kelvin–Helmholtz instability (KHI), the Richardson-number criterion is not satisfied due to the small value of $\eta$. To identify the mechanism responsible for this vertical “puffing up”, we examine the dust distribution and velocities at $t=1000$ $P_{orb}$ in the following subsection. We will also include a comparison between the effective $\alpha$ derived from the vertical dust structure and that inferred from the radial dust distribution.

The resulting radial broadening of the dust distribution prevents $\rho_d/\rho_g$ from increasing indefinitely, implying that additional processes may still be required to raise the dust-to-gas ratio to densities (Hill density) high enough for gravitational collapse. One such mechanism is gas evaporation, which occurs later in the evolution of the disk. We will explore this phase of the disk's evolution and its implications on planetesimal formation in Section~\ref{sec:gas_evap}.

\subsubsection{Vertical versus radial dust profiles}

\begin{figure*}[htbp]
    \centering

    \includegraphics[width=\textwidth]{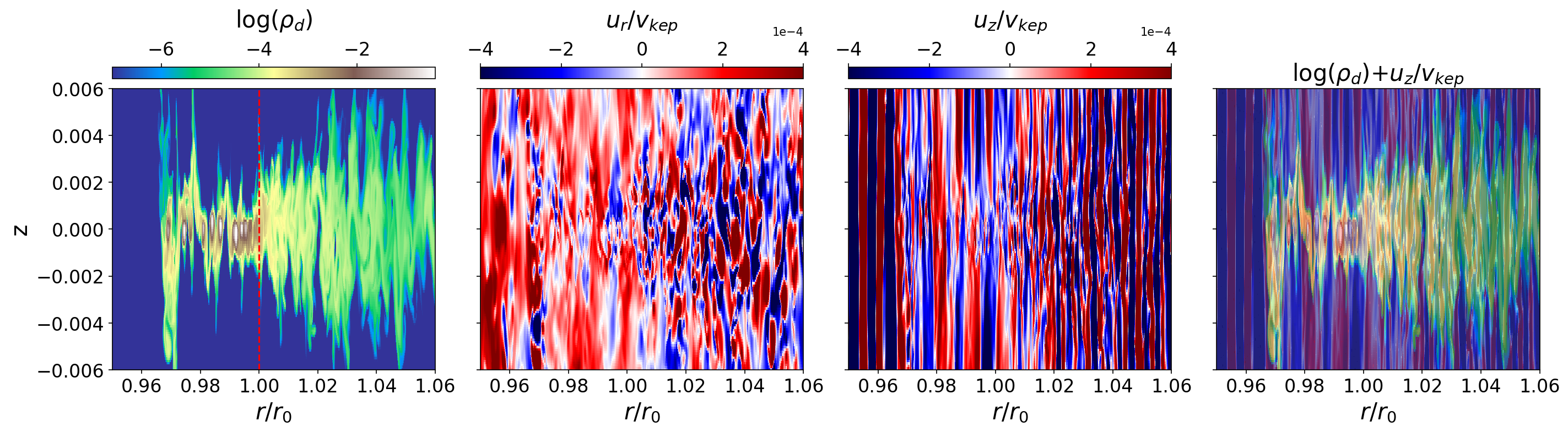}

    \includegraphics[width=\textwidth]{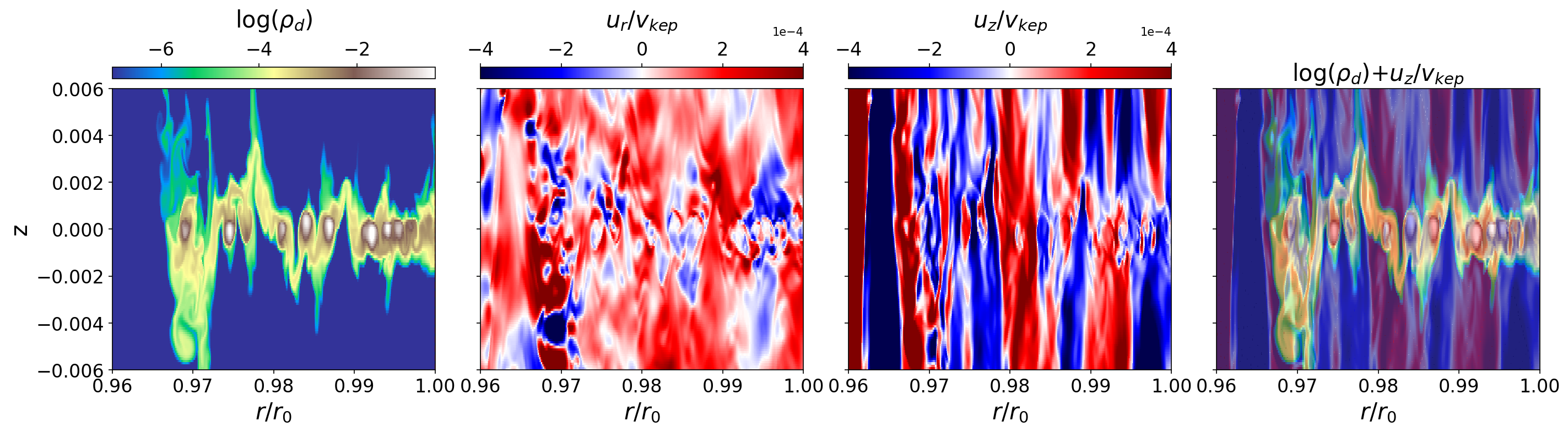}

    \caption{Vertical dust structure as a function of radius for the high-resolution 2D $r-z$ inviscid disk simulation with $\epsilon_0=0.01$ and dust inflow at $t=1000$ $P_{orb}$. The top panel shows a broader radial region, while the bottom panel shows a zoomed-in view of the dust ring region. In the top panel, the outer edge of the dust ring is indicated with a dashed red line. Each set of panels (from left to right) shows: the dust density; the radial gas velocity; the vertical gas velocity (both normalized by the azimuthal Keplerian velocity); and the dust density overlaid with the vertical gas velocity.}
    \label{fig:3D_inviscid_vertical_combined}
\end{figure*}

 The vertical dust distribution and gas dynamics at $t=1000$ $P_{orb}$ are shown in Figure \ref{fig:3D_inviscid_vertical_combined}. As described above, we observe that the vertical dust profile varies with radius, becoming increasingly puffed up outside the main pressure bump. Within the dust rings, the dust settles into a thinner midplane layer than in the surrounding regions, yet still maintains a finite vertical width.  We see alternating patterns of positive and negative vertical gas motions (Fig. \ref{fig:3D_inviscid_vertical_combined} third column), which are reminiscent of the vertical shear instability (VSI). However, we tested that VSI is not responsible for the vertical stirring of the dust. Indeed, in a simulation without dust, the magnitude of the velocities, representing the strength of the VSI, was three orders of magnitude lower than in the simulation with dust.
 
 In addition, an adiabatic simulation with slow cooling produces equivalent results. We think that the strong vertical motion of the gas in the simulation with dust arises because the dust back-reaction, which becomes more significant when there is a high dust-to-gas ratio, tends to compress the gas in the radial direction. Indeed, we observe locations of radial compression in the radial gas velocity profile at the midplane. These compression points can be seen in the second column of Figure \ref{fig:3D_inviscid_vertical_combined}, where the velocity profile transitions from red (outward flowing gas) to blue (inward flowing gas). Because the gas is only mildly compressible by the dust (indeed, the theory of the streaming instability is traditionally developed in the framework of an incompressible fluid), the radial compression of the gas induces an upward vertical flow away from the midplane, as in the streaming instability model \citep{squire2020physical}, entraining and lifting the dust with it. The resulting pattern (Fig. \ref{fig:3D_inviscid_vertical_combined}, third column) consists of alternating red–blue vertical bands, and the spikes in dust density align with the blue (upward flow) regions. The final panel, which overlays dust density on the vertical gas velocity, clearly demonstrates this correlation: dust spikes are co-located with zones of upward-moving gas, coinciding with the locations of strongest radial gas compression (e.g., Fig. \ref{fig:3D_inviscid_vertical_combined} bottom right panel at $r=0.978, 0.99$).

From the vertical dust distribution, we can compute an effective $\alpha$ that induces a sufficient vertical diffusion and compare this to the turbulent $\alpha$ computed from the Reynold's stress tensor $T^{Rey}$. This is usually done by fitting a Gaussian to the vertical dust density profile, from which the vertical height of the dust layer $H_d$ can be extracted, and can then be used to compute $\alpha$ from $H_d = H_g\sqrt{\alpha / \St}\sqrt{\rho_g/(\rho_g+\rho_d)}$. Outside of the dust ring near $r=1.04$ we find $H_d \approx 10^{-3}$ and $\rho_g/(\rho_g+\rho_d) \approx 0.98$, from which we find $\alpha \approx 10^{-5}$. Within the dust ring between $r=0.975-0.985$, we find an average $H_d$ on the order of $\sim 10^{-4}$ and an average $\alpha$ on the order of $\sim 10^{-6}$.

We then compute $\alpha$ from the  $r\phi$ component of Reynold's stress tensor $T^{Rey}_{r\phi}$ using $\alpha=T^{Rey}_{r\phi}/P$, where $P$ is the gas pressure. The value of $\alpha$ is then vertically averaged over the dust-containing region. Near $r=1.04$, we find $\alpha \sim 10^{-5}$, and within the ring, we find an average $\alpha$ on the order of $\sim 10^{-6}$, which are both in agreement with the effective $\alpha$ values computed from the vertical dust height. As such, we can conclude that it is the gas turbulence induced by the dust-back reaction (as it is orders of magnitude smaller if the dust is removed) that limits the settling of the dust.

Finally, we also compute $\alpha$ from the vertical component of the dispersion velocity using the formula $v_z^{rms}=\sqrt{\alpha} c_s$ where $v_z^{rms}$ is the root mean square of the vertical velocity and $c_s$ is the sound speed. Outside the ring we find $\alpha \approx 10^{-4}$, whereas inside the ring we find $\alpha \approx 10^{-5}$, which, curiously, are both an order of magnitude larger than the respective values computed above.

We now turn our attention to the radial width of the ring. The top panel in Fig. \ref{fig:2DRZSigma} shows that the ring is composed of multiple filaments, each at locations of $\eta=0$. An observer with limited resolution would not be able to detect the individual filaments, but would instead interpret this as a single ring with a broad width of 0.025 in $r/r_0$. Interpreting this as a result of turbulent diffusion and using the formula $w_{ring} = \Delta w \sqrt{\alpha_{ring}/\St}$, where $\Delta w$ is the width of the gas profile, one would then conclude that $\alpha_{\mathrm{ring}}\sim 3\times10^{-3}$. Performing the same exercise for the inviscid case without dust inflow (Section \ref{sec:inviscid_noInflow}), and considering the final panel in Figure \ref{fig:3D_inviscid_noInflow_density_snapshots}, we see that the broad width of the dust ring would appear as 0.02 in $r/r_0$, resulting in an $\alpha_{\mathrm{ring}} \sim 2 \times 10^{-3}$.

These values are close to those found in \cite{dullemond2018disk} for several rings observed in protoplanetary disks. This large value of $\alpha$ compared to that obtained from the vertical distribution of dust in protoplanetary disks (Villenave et al., 2022) have prompted the discussion that turbulence might be anisotropic (i.e. stronger in the radial direction than in the vertical direction). Our results suggest that the radial width of rings may not be due to turbulence but to the existence of unresolved filaments separated by a few $10^{-2}$ in relative units. Thus, turbulence in disks does not need to be anisotropic.

\section{Late planetesimal formation (gas evaporation)}\label{sec:gas_evap}

In addressing planetesimal formation, we recall that our code does not include self-gravity of either the gas or the dust. Thus, even reaching Hill density, we would not observe the formation of a self-gravitating clump of dust. Moreover, it has been shown that the real dust dynamics diverge from that of a simulation neglecting self-gravity when the dust density reaches a fraction of a few of $\rho_{\rm Hill}$ \citep{johansen2007rapid}. A simulation without self-gravity, like ours, only gives information about the evolution of the dust-to-gas ratio and not about the absolute values of either of these quantities. This is why in Figures \ref{fig:3D_viscous_density_snapshots}, \ref{fig:3D_viscous_unstrat_density_snapshots}, and \ref{fig:3D_inviscid_noInflow_density_snapshots} we plotted $\rho_d/\rho_b$, where $\rho_b$ is the density that the gas would have in the absence of the pressure bump (Eq. \ref{eq:sigma_gas_background} \& \ref{eq:rho_gas_background}). With a typical gas density at 1 AU of $10^{-9}$ g/cm$^3$, the maximal dust density in Figure \ref{fig:3D_inviscid_noInflow_density_snapshots} would reach $4\times 10^{-7}$ g/cm$^3$, which is the Hill density. However, it is possible that, at a later time in the evolutionary sequence of the disk, the gas was already partially depleted (i.e. $\rho_b \ll 10^{-9}$ g/cm$^3$), or the total amount of dust available (relative to the gas) was smaller than that considered in the simulation. In both these cases, the Hill density would still be out of reach. Thus, here, we consider the late evolutionary phase in which gas is gradually removed through photoevaporation. Under the assumption that at late times all of the available dust has been trapped in multiple rings, gas removal may become the dominant mechanism for enhancing the local dust-to-gas ratio and dust volume density within each ring, offering a possible explanation for the delayed formation of chondrites. We expect gas removal to increase the volume density of the dust by promoting both further sedimentation of dust toward the midplane and radial concentration, for the reasons explained below. We also investigate the possibility of triggering instabilities within the ring which may further promote dust clumping. This will be explored in both a viscous and inviscid disk, in which we remove gas via a uniform exponential decay with a characteristic timescale of 200 orbital periods.

\subsection{Viscous disk}\label{sec:gs_evap_viscous}

We begin by considering a viscous disk, for which we first present a derivation of the expected scaling of the dust density within the ring, $\rho_{d, \rm ring}$, with respect to the dust-to-gas mass ratio $M_d/M_g$. Considering first the vertical profile, the dust scale height is given by 
\begin{equation}\label{eq:dust_height}
    H_d = H_g \sqrt{\frac{\alpha}{\St}\frac{\rho_g}{\rho_g+ \rho_d}}.
\end{equation} 
We also approximate $\rho_g /(\rho_g+\rho_d) \approx\rho_g /\rho_d$, which is valid for $\rho_d\gg\rho_g$ and is the correct regime once enough gas is removed. Then, we can re-write Eq. \ref{eq:dust_height} as
\begin{equation}
   H_d = H_g \sqrt{\frac{\alpha}{\St} \frac{\Sigma_g}{\Sigma_d}\frac{H_d}{H_g}},
\end{equation}
and solving for $H_d$, we find
\begin{equation}\label{eq:dust_scale_height}
    H_d = H_g \frac{\alpha}{\St} \frac{\Sigma_g}{\Sigma_d}.
\end{equation}
Now, we compute the dust ring density at the midplane
\begin{equation}\label{eq:dust_midplane_density}
    \rho_{d, \rm ring, 0}=\frac{\Sigma_d}{\sqrt{2\pi}H_d} =\frac{\Sigma_d}{\sqrt{2\pi (\alpha/\St)}H_g} \left[ \frac{1}{\sqrt{\alpha/\St}} \frac{\Sigma_d}{\Sigma_g}\right].
\end{equation}
The first fraction term is the midplane density in the regime of low dust-to-gas ratio and the second term inside the brackets is the enhancement factor, which is proportional to $\Sigma_d/\Sigma_g$. Formulae \ref{eq:dust_scale_height} and \ref{eq:dust_midplane_density} show that $H_d$ and $\rho_d$ depend on $\Sigma_d/\Sigma_g$, which increases as gas is removed. The reason is that the effective $\alpha$ is $\alpha[\rho_g/(\rho_d+\rho_g)]$. If we had not considered the quenching of alpha by the dust-to-gas ratio, $H_d$ and $\rho_d$ would not have appeared to change during gas removal, which is unphysical (in absence of gas there is no reason for which the dust should have a scale height). We stress that in our simulations we will keep $\St$ constant  during gas removal. In reality, the dust velocity dispersal also scales as $\sqrt{\alpha[\rho_g/(\rho_d+\rho_g)]\St} \:c_s$, so that, in a fragmentation limited regime, $\St$ should increase as $\rho_g/(\rho_d+\rho_g)$ decreases. This enhances dust sedimentation.

Considering the radial profile, the equation for $\Sigma_{d}$ at the center of the dust ring is given by $2\pi r \Sigma_d = M_d/(\sqrt{2\pi} w_d)$, where $M_d$ is the mass of the dust in the ring and $w_d$ is its Gaussian width. The same is true for the gas, and is given by $2\pi r \Sigma_g = M_g/(\sqrt{2\pi} w_g)$. At the center of the ring, we then have 
\begin{equation}\label{eq:sigmaRatio_ring_center}
    \frac{\Sigma_d}{\Sigma_g} = \frac{M_d}{M_g}\frac{w_g}{w_d}.
\end{equation}
Plugging this into Eq.~\ref{eq:dust_midplane_density}, we obtain the dependence of the dust density on the mass ratio. The radial contraction of the dust ring width is also expected to depend on $M_d/M_g$, following the same approach presented for the vertical contraction; however, there are two important differences between radial contraction and midplane settling. The first is that, the locations at which $\eta=0$ are not in steady state and shift radially over time, so the dust continuously readjusts to follow these moving trapping points. Instead the midplane obviously does not move. The second is that, while the settling timescale is proportional to $1/(\Omega \St)$, the radial contraction timescale is proportional to $\delta r/(\St \eta v_k)$, where $\delta r$ is the distance from the pressure bump. The factor $\delta r /\eta \gg 1$, so the contraction timescale is much longer than the sedimentation one. The gas may be removed faster than the dust has the time to radially adjust. In summary, although some degree of radial contraction is still expected, this time-dependent structure prevents a simple analytical description of the ring width. As a result, while the midplane dust density in the ring is expected to scale as a power of ($M_d/M_g$), the exact exponent cannot be determined without the expected scaling for the radial width evolution. 

To investigate this behavior, we initiate the gas-removal phase from an intermediate stage of our previous 3D stratified viscous disk (Section~\ref{sec:3D_viscous_stratified}), by which time a well-developed dust ring has already formed. We also repeat the same gas evaporation simulation in a 2D $r-z$ disk at identical resolution, which provides a controlled comparison to isolate the role of the azimuthal dimension and to highlight additional effects that may arise in the full 3D evolution, such as clumping.

\begin{figure}[htbp]
    \centering
    \includegraphics[width=0.48\textwidth]{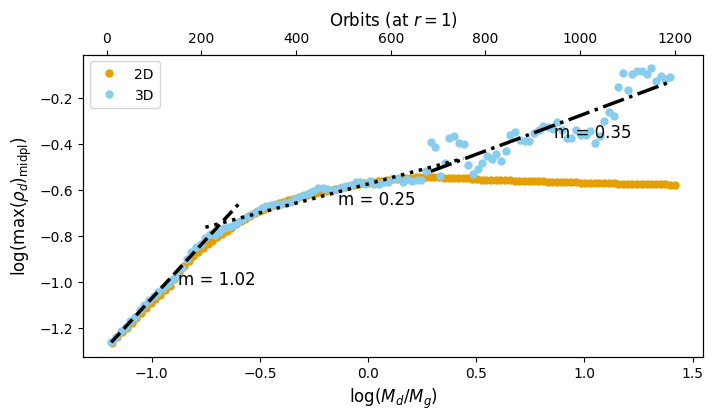}
    \caption{Maximum dust density in the ring as a function of the dust-to-gas mass ratio during gas removal for both a 2D (orange) and 3D (blue) viscous disk ($\alpha=10^{-4}$). The upper x-axis indicates orbital periods at the reference radius $r=1$. Linear best-fit lines are shown for three stages of the gas evaporation process, with their slopes indicated next to each line. The slopes of the first two stages are consistent between the 2D and 3D case, and then diverge in the final stage during which there is clumping in 3D.}
    \label{fig:3D_gasEvap_visc_full}
\end{figure}

Figure~\ref{fig:3D_gasEvap_visc_full} shows the evolution of the maximum midplane dust density within the ring, plotted as a function of the total dust-to-gas mass ratio in the simulation domain for both the 2D (orange) and 3D (blue) cases. The evolution can be divided into three distinct stages, each characterized by a different scaling. The 2D and 3D evolution closely follow each other during the first two stages and diverge only in the final stage. In the first stage, with a slope of $m=1.02$ (dashed line), gas removal enhances dust sedimentation towards the midplane (Eq. \ref{eq:dust_scale_height}) while the radial width of the ring also contracts. Radial contraction contributes a fraction of approximately 0.4 of the density increase, while vertical settling accounts for the remaining 0.6. The latter is less than expected from Eq. \ref{eq:dust_midplane_density}, but this is because we are already close to vertical resolution. Indeed, in the second stage, the thickness of the dust layer becomes limited by the vertical resolution, preventing further settling, so density growth is driven only by radial contraction, resulting in a weaker scaling of 
$m=0.25$ (dotted line). During this phase, sub-rings also begin to form within the main dust ring.

In the final stage, the 2D and 3D cases clearly diverge. The dust ring develops additional sub-structures, forming two distinct sub-rings, and radial contraction effectively ceases. This is clear in the 2D case, where we see that the dust density saturates. In contrast, the 3D simulation shows renewed growth, with a steeper scaling of $m=0.35$ (dash-dotted line) This enhancement is driven by the development of azimuthal asymmetries within the rings, indicating the formation of dust clumps, a process that is absent in the 2D case. Once sufficient gas has been removed, the dust forms clumps that continue to grow in density, such that we expect the Hill density to eventually be reached and gravitational collapse to occur.

\subsection{Inviscid disk}\label{sec:gas_evap_inviscid}

\begin{figure*}[htbp]
    \centering
    \includegraphics[width=\textwidth]{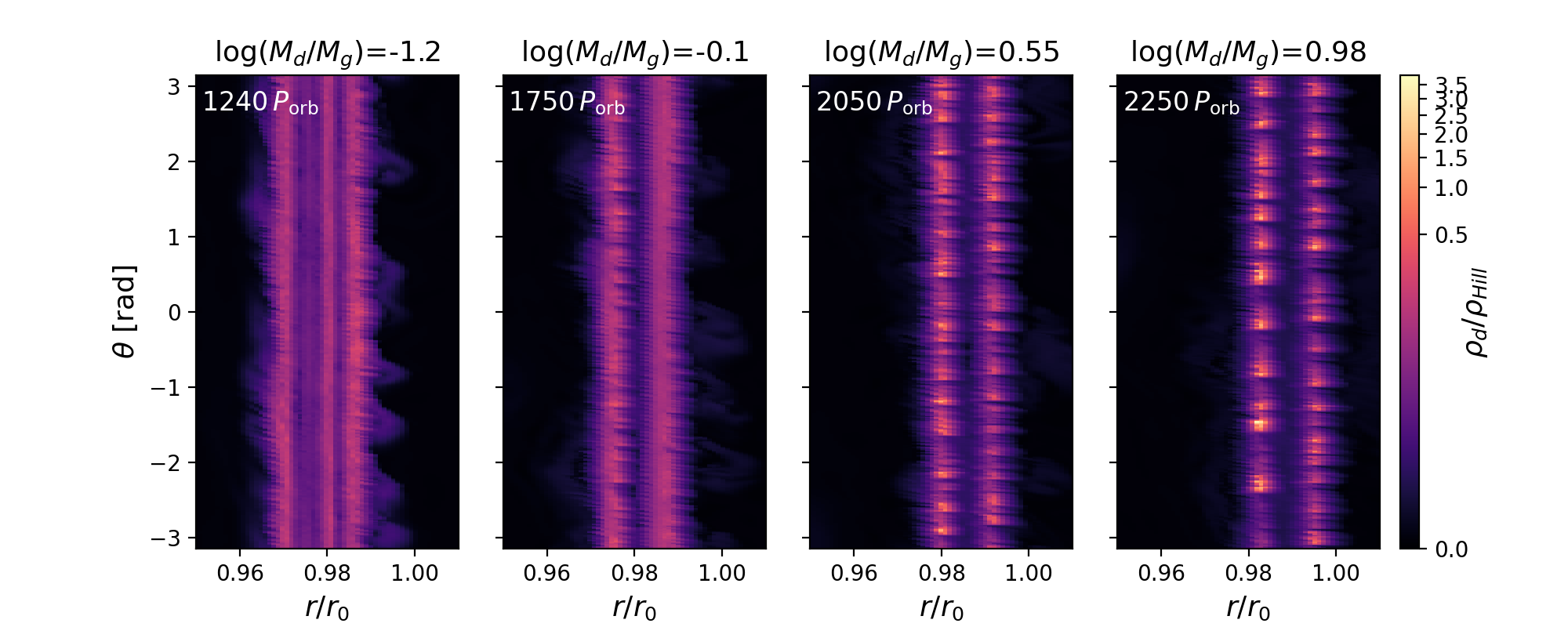}
    \caption{Similar to Fig. \ref{fig:3D_inviscid_noInflow_density_snapshots}, but showing the evolution of the dust density as gas is removed from the disk. The dust density is plotted with respect to the Hill density and the total dust-to-gas mass ratio at each given time is indicated at the top of the panel. The orbital period in the top left corner is given relative to the start of the original inviscid run with no inflow in Section~\ref{sec:inviscid_noInflow}.}
    \label{fig:3D_noVisc_gasEvap_density_snapshots}
\end{figure*}

We next investigate gas evaporation in an inviscid disk. In this case, we initiate the gas-removal phase from the previous inviscid run without dust inflow (Section~\ref{sec:inviscid_noInflow}) at $t=1240 P_{orb}$, at which point the disk already exhibits well-developed dust ring substructures, but no clumps. Our analysis of the subsequent evolution of the dust density focuses on the dominant sub-ring, i.e., the most massive of the two principal sub-rings that eventually form in Figure \ref{fig:3D_inviscid_noInflow_density_snapshots}. 

Figure \ref{fig:3D_noVisc_gasEvap_density_snapshots} shows the evolution of the dust rings as gas is gradually removed. Within the ring, the thickness of the dust layer is already limited by the vertical resolution, so initially the growth of dust density is due only to radial contraction. Considering the dominant inner sub-ring, we find that the ring width $w_d$ contracts as $w_d \propto (M_d/M_g)^{-0.19}$, until we reach $\log(M_d/M_g)\approx 0.4 - 0.5$, after which further contraction is limited by the radial resolution.

Eventually, we also observe the formation of dust clumps. To quantify the growth rate of the clumping, we compute the ratio, at the midplane, of the maximum dust density within the ring to the azimuthally averaged dust density at the center of the ring. We find that this ratio scales as $(M_d/M_g)^{0.5}$, indicating that there are azimuthal asymmetries within the ring that grow faster than the average dust density, and thus signaling the formation of dust clumps. Note that clumps have not formed at the corresponding times in the simulation shown in Fig. \ref{fig:3D_inviscid_noInflow_density_snapshots}, indicating that gas removal is primarily responsible for dust clumping. These clumps would eventually reach Hill density and, as in the viscous case, are therefore expected to undergo gravitational collapse.

In summary, we observe dust clumping in both viscous and inviscid cases, with the maximum dust density growing as $(M_d/M_g)^{0.35}$ and $(M_d/M_g)^{0.5}$, respectively. In the viscous case, the late-stage evolution becomes similar to that of the inviscid case as gas is progressively removed. This convergence arises because gas depletion drives the disk toward the inviscid limit: even for a fixed $\alpha$-viscosity, the effective dust diffusivity is reduced by the factor $\rho_g/(\rho_g+\rho_d)$. In both cases, the dust clumps are therefore expected to undergo gravitational collapse.

\section{Conclusions}\label{sec:conclusions}
In this work, we first reproduced and extended the 2D shearing-box results of \cite{liu2023dusty} in a global 2D disk simulation. In a relatively mild pressure bump with the same parameters as LB23, we recover the Type II DRWI and the associated formation of dust clumps. By continuously supplying dust to the ring, we extend their analysis by characterizing the evolution of these clumps, finding an average of 20–25 clumps and reaching a maximum dust-to-gas surface density ratio of $\Sigma_d/\Sigma_b \approx 34$.

In contrast, we find that the DRWI is suppressed in fully 3D disks with the same viscosity (here $\alpha=10^{-4}$ as in LB23): no dust clump develops and the dust maintains a Gaussian distribution in $r$ and $z$ around the ring center. We attribute this suppression to dust sedimentation, which concentrates solids at the midplane while leaving the upper and lower layers dust-poor and dynamically stable. Viscous coupling between vertical layers then acts to damp perturbations that arise at the midplane. This interpretation is supported by the unstratified 3D simulation, in which dust is present at all heights and the disk exhibits stronger perturbations reminiscent of the 2D behavior.

In inviscid 3D disks, we find that the dust ring evolves differently. Multiple sub-rings form because the back-reaction of the dust on the gas changes the gas density enough to form new pressure maxima around the original (single) maximum. When dust is continuously supplied, instead of enhancing the dust density in the already existing rings, it modifies the overall gas density profile, partially flattening and shifting the global pressure bump. This allows for the formation of more numerous sub-rings, over a more extended radial region. As a result, the dust density may remain everywhere lower than the threshold for gravitational instability. We stress that our code does not account for self-gravity, so the results are independent of the gas or dust densities but dependent only on their ratio. Thus, whether Hill density is reached depends on the initial assumed density. 

This result reveals that the radial width of the dust rings imaged by ALMA in protoplanetary disks can arise from the presence of multiple, closely spaced sub-rings that remain unresolved at observational resolutions. When interpreted as a single broad ring, the contrast with the vertical thinness of the dust layer suggests a much larger turbulent diffusion in the radial direction than in the vertical direction \citep{dullemond2018disk}. Therefore, our results suggest that this inferred anisotropy of turbulence may not be real.

Finally, we investigate the late evolutionary stages of disks by modeling gas loss, while noting that a detailed photoevaporation model is not included. In inviscid disks undergoing gas depletion, we find growing azimuthal asymmetries within the dust rings, consistent with the formation of dust clumps. In viscous disks, the evolution eventually resembles the inviscid case, including clump formation, once sufficient gas has been removed; as gas is depleted, the effective $\alpha$ in the dust diffusion coefficient is reduced and thus the disk approaches the inviscid limit. 

Overall, our results suggest that planetesimal formation within pressure bumps is favored in very low-viscosity disks and can be triggered during late stages of disk evolution, when sufficient gas has been removed by photoevaporation. The amount of gas removal required depends on the gas viscosity. This picture is consistent with the inferred late formation times of the parent bodies of chondritic meteorites in the Solar System.

\begin{acknowledgements}
  AM and MT are grateful for support from the ERC advanced grant HolyEarth N. 101019380. This work was granted access to the HPC resources of IDRIS and CINES under the allocation  A0180407233  made by GENCI. EL and MT wish to thank Alain Miniussi for maintenance and re-factorization of the code fargOCA.
\end{acknowledgements}

\newpage
\bibliographystyle{aa} %
\bibliography{DRWI}

\end{document}